\documentclass[aip, amssymb,showpacs,supersriptaddress,groupeaddress]{revtex4-2}
\usepackage{amssymb}
\usepackage[percent]{overpic}
\usepackage{graphicx}
\usepackage{mwe}
\usepackage{braket}
\usepackage{mathtools, amssymb, amsmath}
\usepackage{cancel}
\usepackage{float}
\usepackage{mathrsfs}
\usepackage{gensymb}
\usepackage[utf8]{inputenc}
\usepackage{amsmath}
\usepackage{graphicx}
\usepackage{dcolumn}
\usepackage{bm}
\usepackage[title]{appendix}
\usepackage{subcaption}
\usepackage{bm}
\usepackage{natbib,hyperref}
\usepackage{hyperref}
\usepackage{color}
\usepackage[dvipsnames]{xcolor}
\usepackage[normalem]{ulem} 

\hypersetup{colorlinks=true, 
    linkcolor=red,          
    citecolor=magenta,        
    filecolor=gree,      
    urlcolor=blue           
}

\usepackage{float}

\usepackage[utf8]{inputenc}
\usepackage[T1]{fontenc}
\usepackage{mathptmx}
\usepackage{etoolbox}
\usepackage{xr-hyper}
\usepackage{hyperref}

\begin{document}
\title{Synaptic plasticity in Co/Nb:STO memristive devices: the role of oxygen vacancies}

\author{Walter Quiñonez}
\affiliation{Instituto de Nanociencia y Nanotecnología (INN), CONICET-CNEA-CAC, 1650, Buenos Aires, Argentina.}

\author{Anouk Goossens}
\affiliation{Zernike Institute for Advanced Materials, University of Groningen, 9747 AG, Groningen, The Netherlands.}
\affiliation{Groningen Cognitive Systems and Materials Center,
University of Groningen, 9747 AG, Groningen, The Netherlands.}
\author{Diego Rubi} 
\affiliation{Instituto de Nanociencia y Nanotecnología (INN), CONICET-CNEA-CAC, 1650, Buenos Aires, Argentina.}
\author{Tamalika Banerjee}
\affiliation{Zernike Institute for Advanced Materials, University of Groningen, 9747 AG, Groningen, The Netherlands.}
\affiliation{Groningen Cognitive Systems and Materials Center,
University of Groningen, 9747 AG, Groningen, The Netherlands.}
\author {María José Sánchez} 
\affiliation{Centro At\'omico Bariloche and Instituto Balseiro (UNCuyo), 8400 San Carlos de Bariloche, R\'io Negro, Argentina.}
\affiliation{Instituto de Nanociencia y Nanotecnolog\'{\i}a (INN),CONICET-CNEA-CAB, 8400, San Carlos de Bariloche, Argentina.}
\email[Corresponding author:]{maria.sanchez@ib.edu.ar}

\date{\today}

\begin{abstract}

Neuromorphic computing aims to develop energy-efficient devices that mimic biological synapses. One promising approach involves memristive devices, that can dynamically adjust their electrical resistance in response to stimuli, similar to synaptic weight changes in the brain.
However, a key challenge is understanding and controlling the coexistence of different types of synaptic plasticity, like short-term and long-term plasticity.
In this work, we show that plasticity behaviors in Co/Nb:STO Schottky memristors originate from oxygen vacancy electromigration, which modulates the Schottky barrier and enables both short-term and long-term plasticity. Our experiments  reveal that resistance changes follow a power-law during reading (short-term plasticity) and increases stepwise with successive pulses (long-term memory retention). These behaviors are successfully reproduced by our model that  demonstrates the correlation between oxygen vacancy distribution and Schottky barrier modulation. Our findings highlight these memristors as promising candidates for neuromorphic applications.

\end{abstract}
\maketitle

\section{Introduction}

The drive to overcome the limitations imposed by the von Neumann bottleneck has spurred 
the development of advanced materials and devices that transcend conventional silicon technologies \cite{najmaei_2022, park_2023}. Neuromorphic computing has emerged as a promising paradigm designed to achieve more efficient and brain-inspired information processing \cite{upadhyay_2019, kendall_2020, zhang_2020}. The principal requisites for neuromorphic devices can be broadly categorized into energy efficiency, compact integration, and brain-like functionalities, which are further divided into neuronal and synaptic functionalities to emulate different aspects of brain-like processing.

Among the various technologies underpinning neuromorphic computing, memristors \cite{val_2013, kumar_2022} have emerged as key enablers for
hardware implementation of such architectures \cite{li_2018,mehonic_2020, sebastian_2020, yao_2020, sun_2021, zhou_2022}.
Memristors are metal/insulator/metal capacitor-like structures that exhibit resistive switching (RS), where electrical resistance changes reversibly in response to applied electrical stimuli
\cite{saw_2008,yan_2008,waser_2010,borghetti_2010, pan_2014,wang_2020}.

Functional oxides are particularly well suited materials to built memristive devices due to their highly tunable physical and electrical properties, responsiveness to external stimuli and promising compatibility with modern
CMOS integration \cite{park_2023}.
Consequently, oxides-based memristors \cite{waser_2010,lee_2011, breuer_2016,kumar_2022,xiao_2023,patil_2023} are valued due to their highly tunable physical and electrical properties, low power consumption, potential scalability and their ability to exhibit non-volatile and/or volatile RS.


In non-volatile (NV) RS, resistive states persist even after the electrical stimulus is removed. In particular, memristors that exhibit NV RS are promising candidates for future memory devices due to {their  
simple structure, fast switching time, high density integration, low power consumption, retention times and 3D stacking possibility \cite{was_2009,waser_2010}.
In contrast, volatile RS (VRS) exhibits resistive states that spontaneously relax once the electrical stimulus is turned off \cite{zhuo_2013, zhou_2022}. 
NV memristors can effectively mimic biological long-term plasticity, enabling the construction of artificial neural networks using crossbar arrays \cite{perez_2019,xia_2019}. 
VRS, on the other hand, can emulate short-term plasticity, which plays a crucial role in neural network dynamics and is particularly well-suited for reservoir computing systems \cite{iel_2018}.
A variety of synaptic plasticity mechanisms have been implemented in hardware, including long-term and short-term potentiation and depression 
\cite{iel_2019, zhuo_2013,zhou_2022} as well as spike-timing-dependent plasticity (STDP), among others. 
These characteristics naturally facilitate  "in-memory" computation \cite{wu_2023}, mirroring the adaptive synaptic strengths of neural networks, which must be dynamically updated as learning occurs.



The modulation of the Schottky barrier formed at the metal-insulator interface due to the dynamics of oxygen vacancies (OV) is one of the most widely accepted mechanism for interfacial RS in several oxide-based memristors \cite{saw_2008, roz_2010, iel_2016,valov_2017,pan_2014}. 
In particular, metal/Nb:SrTiO$_3$ (STO) Schottky junctions have been extensively investigated and several works show that the emergence of RS is governed by 
the quality of the metal/STO interface \cite{pan_1_2014,mikheev_2014}.

We have demonstrated in earlier works \cite{Goossens_2023,Goossens_2018_APL,Goossens_2023_APL} interface Schottky memristors using Co electrodes on Nb:STO. Such interfaces exploit the non-linear variation of the dielectric permittivity in Nb:STO with electric field and exhibit multi-level switching with large ON/OFF ratios and minimal device-to-device and cycle-to-cycle variations \cite{Goossens_2023,Goossens_2018_APL}. In addition, cross-sectional scanning transmission electron microscopy (STEM) combined with energy-dispersive X-ray spectroscopy (EDS)  \cite{Goossens_2023}, reveal  the presence of an homogeneous oxygen-deficient layer at the interface which allows for the trapping and detrapping of carriers. By performing an electrostatic analysis \cite{Goossens_2023_APL}, a DR length of approximately 15 nm was estimated for Co/Nb:STO devices with a doping concentration of 0.1 wt$\%$. 
We have also studied how memristive characteristics such as ON/OFF ratio, stochasticity and nonlinearity are influenced by varying the doping of the Nb:STO semiconductor \cite{Goossens_2023_APL}, opening prospects in applications such as data encryption and random number generation.

In this study, we fabricated such two-terminal Co/Nb:STO memristors, with the primary goal of exploring their capability to mimic synaptic plasticity.
We specifically focus on the potentiation of resistive states under electrical pulsing schemes with varying parameters. Our experiments revealed a combination of NVRS and VRS behaviors, emulating an intertwined effect of LTP and STP.
Furthermore, we observed that these processes are governed by power laws with distinct exponents. Based on these findings, we propose a theoretical model that accounts for  OV electromigration, demonstrating how the movement of OV modulates the Schottky barrier over time, thereby reproducing the electrical response observed in the experiments.
This work provides valuable insights into the mechanisms underlying synaptic behavior in Co/Nb:STO memristors, attributing it to OV electromigration rather than conventional models of RS based on charge trapping and detrapping, where defects are treated as fixed traps within the material.

\section{Methods}
For the fabrication of the devices (see Fig.\ref{Fig.device}(a)),  Nb:STO (001) substrates with doping concentration of 0.1 wt$\%$ from Crystec was used. The substrate was chemically treated with buffered hydrofluoric acid (BHF) and subsequently annealed at 960$^\circ\text{C}$ in an O$_2$ flow of 300 ccmin$^{-1}$. 
Electron beam lithography and electron beam evaporation were used to fabricate circular Co electrodes (20 nm thick) with a radius of 2 $\mu$m. The electrodes were electrically isolated by an insulating AlO$_x$ layer to minimize device crosstalk. To prevent oxidation and protect the contacts, a 100 nm Au capping layer was deposited. Further details on the fabrication process can be found in \cite{Goossens_2023}. The chip was mounted onto a carrier using Ag paste, which simultaneously functioned as the back electrode. Electrical characterization was conducted using a Keithley 2450 SMU connected to a probe station.

\begin{figure}[H]
\centering
\includegraphics[width=1 \textwidth]{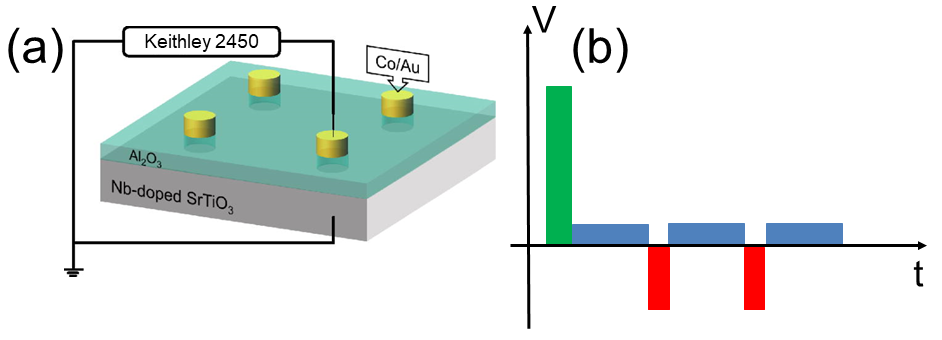}
\caption{(a) Schematic of the fabricated devices. Circular Co electrodes were deposited on top of the Nb-doped SrTiO$_3$ substrate to generate the active layer. A capping layer of Au was deposited above the Co electrodes, forming an ohmic interface. An insulating layer of AlO$_x$ was deposited to prevent crosstalk. The electrical forward-bias connection was achieved by connecting the positive terminal of the SMU to the Au layer and the negative terminal to the Nb:STO using Ag paste. (b) Electrical protocol used for the measurements. The green pulse represents large positive voltages to set de device in a low resistive state, while the blue pulse is  a low constant bias used to read the resistive state. Red pulses represent write voltages use to gradually increase the resistive state. }
\label{Fig.device}
\end{figure}

\section{Results}

\begin{figure}[H]
\centering
\includegraphics[width=1 \textwidth]{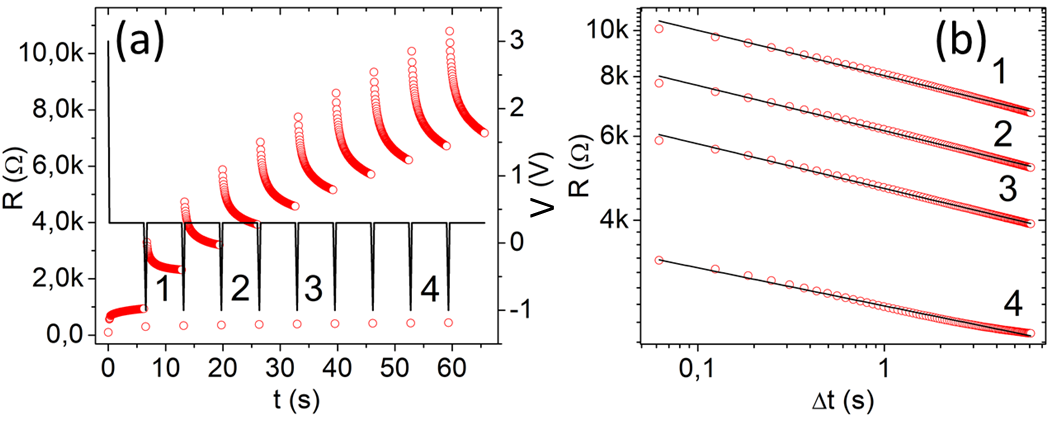}
\caption{Double time scale in the writing pulse accumulation and read process. (a) Evolution of the resistance (left vertical axis, red dots) and applied voltage V (right vertical axis, black solid line) vs time. 
(b) Experimental resistance (red dots) vs time interval $\Delta t$- measured after the corresponding writing pulse for the reading process 1,2,3,4 indicated in (a). Each curve was fitted by a power law (black line). See table I for details.}
\label{Fig.resul1}
\end{figure}

To emulate the synaptic response of the device, the voltage source acted as the presynaptic neuron and was connected in forward bias to the Co electrode of the device. The spiking activity of the presynaptic neuron was emulated by a pulsed scheme in which pulses higher than -0.8 V are considered to be the writing voltages, while a constant bias of +0.3 V-referred to as the reading bias- is used to register the resistance of the device (see Fig.\ref{Fig.device} (b) for a scheme of the pulsing protocol). 
The protocol started with a +3 V pulse to set the device in low resistance state (LRS). Subsequently a reading bias was applied for $\sim$6 s, followed by a writing pulse of -1 V applied for 62 ms, after which, the reading bias was applied again. 
This pattern of write/read processes was repeated 9 times. In Fig.\ref{Fig.resul1}(a) it can be seen that after each writing pulse the resistance increases abruptly relative to the previous resistance state.
This suggests a cumulative effect in the modulation of the device’s resistance, analogous to synaptic weight strengthening in biological neurons, commonly referred to as long-term potentiation (LTP). 

Conversely, during the application of the reading bias, a gradual decrease in resistance was observed, indicating a volatile memory effect where the stored information partially dissipates over time in the absence of further stimuli. This resembles short-term synaptic plasticity (STP), where memory fades unless reinforced by subsequent neuronal activity.

In Fig.\ref{Fig.resul1}(b) the resistances measured during each of the reading periods $\Delta t$, corresponding to time intervals labeled 1, 2, 3, and 4 respectively in Fig. \ref{Fig.resul1}(a), are extracted and plotted on a log-log scale. 
Each dataset was fitted to a power law function $R = a{\Delta t}^b$ with the parameters obtained from the fits presented in Table \ref{tab:parameters}.
Notably, the exponent $b$ remained consistent across all intervals, indicating no significant variation as writing pulses accumulated over time. 

To explore how the envelope of the R vs t curves can be modulated, the pulse protocol was varied, firstly by modifying the duration of the reading bias. In Fig.\ref{Fig.result2}(a) the resistance vs time is shown for 870 ms (black dots), 1.5 s (magenta dots), 3 s (red dots) and 6 s (blue dots) 
, while keeping the reading bias (+300 mV) and writing pulse (-1 V) unchanged.
To facilitate comparison across different reading periods, Fig.\ref{Fig.result2}(b) presents the resistance values (normalized to the initial value) measured for the first read after each writing pulse, following the standard LTP 
measurement protocols. 

\begin{table}[ht]
   \centering
    \begin{tabular}{|c|c|c|}
         \hline
            & a [Ohm] &  b [1]\\
         \hline
         Curve 1& 8041.73 & -0.095 \\ 
         \hline
         Curve 2  & 6163.3 & -0.095  \\
         \hline
        Curve 3& 4665.17 & -0.093  \\ 
         \hline
         Curve 4 & 2646.42 & -0.079 \\
         \hline
    \end{tabular}
    \caption{Parameters a and b obtained from fitting the curves labeled 1-4  of Fig.\ref{Fig.resul1}(b) with a power law $R = a\Delta t^b$.} 
    \label{tab:parameters} 
\end{table}

This allowed us to isolate the envelope curve for each reading interval, effectively shifting the analysis from the time domain to the number of accumulated pulses. We observe an almost perfect overlap of all the envelope curves, indicating that this resistance evolution is dominated by the non-volatile effect (LTP). We notice, in addition, that a single reading pulse -recall the protocol used to obtain Fig. \ref{Fig.resul1}(b)- does not seem to be enough to trigger a noticeable volatile effect, in the range of explored time intervals $\Delta$t. This confirms that the volatile effect (STP) requires the accumulation of multiple reading pulses, as seen in Fig. \ref{Fig.resul1}(a). 

We also analyzed the impact of varying the writing pulse magnitude while keeping the reading time (3 s) and reading voltage (+300 mV) fixed. Fig. \ref{Fig.result2}(c) shows resistance evolution for writing pulses of -0.8 V (red dots), -1 V (black dots), -1.3 V (blue dots), and -1.5 V (magenta dots).
In this case, it is evident that higher writing voltages result in significant gradual changes in the resistance.
This is highlighted in Fig.\ref{Fig.result2}(d), where we obtain four different power laws when plotting the envelope curve of the resistance (normalized to the initial value) as a function of the accumulated number of pulses.  

\begin{figure}[H]
\centering
\includegraphics[width=1 \textwidth]{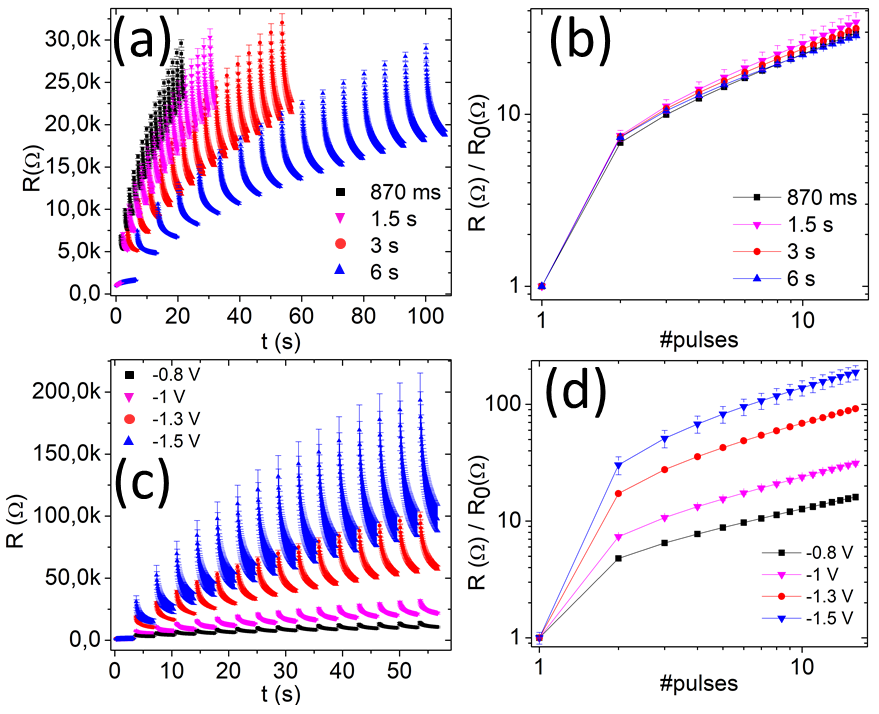}
\caption{ Variation of the reading time window and the writing voltage. Data are as a function of time in (a) and as a function of the number of writing pulses in (b). Changing the time window for the reading process does not change the shape of the envelope, as all the curves collapse to the same in (b). 
When the intensity of the writing voltage changes, going from the time (c) to the number of writing pulses (d) domain, results in different power laws for each writing voltage.} 
\label{Fig.result2}
\end{figure}

\begin{figure}[H]
\centering
\includegraphics[width=1 \textwidth]{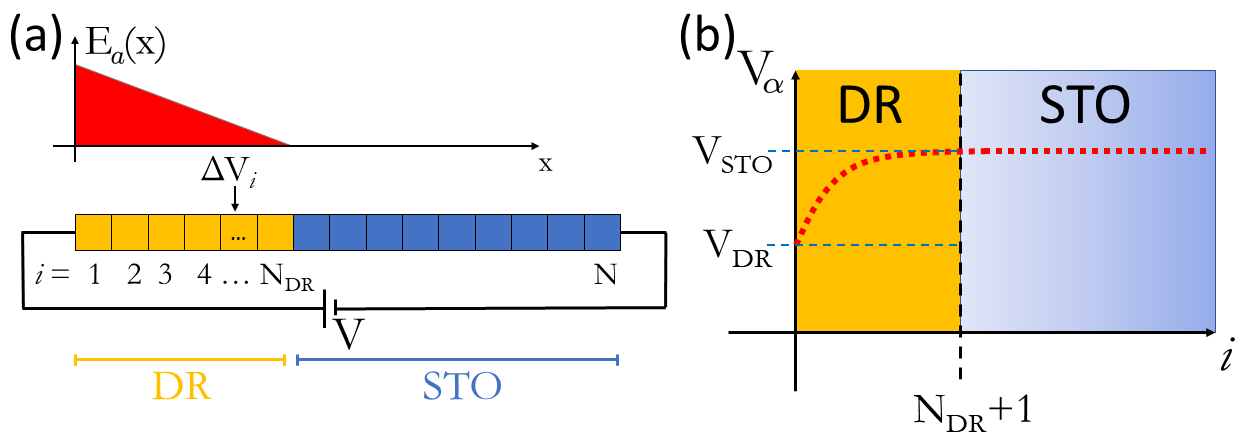}
\caption{(a) 1D model of the device interface. From a total of N = 100 sites, we assign N$_{DR} = 25$ to the DR, while the others correspond to the STO. We assume a linear drop of the applied electric field $E_a$.
(b) Electric potential  profile $V_{\alpha}$: For the STO a constant activation energy is taken, but in the DR the activation energy is a function of site $i$ and takes into account the field generated by the fixed donors in the DR. 
}
\label{fig:Model}
\end{figure}

To model the experimental response of the device, we 
build upon the well-established relationship between resistance modulation, the voltage drop across a Schottky interface, and the density of trap states in the depletion region \cite{zhao_2019,yin2015role,fan2017resistive,bourim2014interface}. The resistance of the junction is mainly determined by the Schottky barrier whose height depends on the defect concentration \cite{mikheev_2014}.
 Additionally, OV are ubiquitous 
 defects present in metal-oxide interfaces that can act as carriers trapping/detrapping centers \cite{gritsenko_2016,zhao_2019,ni_2021}.
 In particular, the modulation of the barrier height 
 induced by OV migration has been recently suggested in CoO/Nb:STO interfaces \cite{kunwar_2023}. In addition,  in our previous  work \cite{Goossens_2023}, STEM-EDS analysis revealed that there is no significant OV clustering in the bulk of the STO, nor evidence of conductive filament formation, indicating that the resistive changes occur at the Co/Nb:STO interface.

 With these ingredients at hand, we 
 consider OV as traps present at the interface between the Co layer and the Nb:STO and study the OV dynamics under 
 an applied voltage across the Schottky interface 
 to explain the observed resistance changes.
 To model this behavior, we represent the interface as a 1D chain with $N = N_{DR} + N_{STO}$ sites, where $N_{DR}$ 
 correspond to the depletion region (DR) and $N_{STO}$ 
 to the Nb:STO region (STO) outside of the DR (see Fig.\ref{fig:Model} (a)). Each site $i$ contains a number $n_i(t)$ of OV at time $t$ and we define the fraction (or concentration) of OV as 
$\delta_{i} (t)= n_i(t)/n_T$, 
where $n_T$ is the total number of OV present 
across the whole interface.
We assume that the STO outside the DR acts as a reservoir where OV can accumulate without substantially affecting the STO's resistivity, in agreement with previous reports \cite{desouza_2012,trabelsi_2017}.

As positively  charged defects, OV can migrate from the DR into the STO region 
under a positive applied stimulus (voltage) and, if the polarity of external voltage is reversed, return to the DR. 
Consequently, the total number of traps (OV) in the DR increases or decreases over time, directly influencing the device's resistance.

Following the ideas of the 1D VEOV model, originally introduced in Ref.\onlinecite{roz_2010}, for a given external voltage applied at time $t$, at each simulation step the fraction of OV per site was updated with the rate probability (probability per unit time):
\begin{equation}
p_{ij} = \delta_i (1-\delta_j) \exp(-V_{\alpha} + \Delta V_{i}), 
\label{pij}
\end{equation}
for a transfer from site \textit{i} to a nearest neighbor \textit{j}= \textit{i} $ \pm 1$. 
Note that $p_{ij}$ is proportional to the concentration of OV at a given site and to the available concentration at the neighboring sites.
In the Arrhenius factor, $\exp(-V_{\alpha} + \Delta V_{i})$,
$V_{\alpha}$ includes the contribution of the activation energy for pure OV diffusion  (in the absence of an external voltage), as well as the contribution to the potential energy 
from fixed donors (others than OV) as a function of the position $i$ in the DR. Outside the DR, we only consider a constant activation energy for OV diffusion (see Fig.\ref{fig:Model} (b))
The factor $\Delta V_i$, on the other hand, comprises the contributions from the applied external voltage and 
the local field generated by the fraction of OV at each site (see Supplementary Material). All the energy scales were taken in units of the thermal energy $K T$.

Using Eq.(\ref{pij}),  we updated the OV concentration at each time step and computed, for instance,  the total charge due to OV along the DR, $Q(t) = q \,n_{DR}(t)$, where $q$ is the charge of a single OV and 
 $n_{DR}(t) \equiv \sum_{i=1}^{N_{DR}} n_i (t)$ is 
the total number of  traps (OV) along the DR. For more details, see the Supp. Material where we included a flowchart of the model.

The height of the Schottky barrier was computed using the well-known relation $\Phi_b (t)= W + Q(t) \, d/\epsilon$ \cite{mikheev_2014}, where $W $ includes the difference between the metal M work function and the electron affinity of the STO and an extra contribution that might come from fixed donors. In the last term, $d$ is the length of the DR and $\epsilon$ is the dielectric permittivity of STO, which, for simplicity, we assume to be constant for a given applied voltage. 

For forward bias $V(t)$, the current (density) was computed using the field-enhanced thermionic emission equation \cite{sze_1981} which 
governs the transport of carriers over a Schottky barrier at room temperature:
\begin{equation}
  I_s(t)= A T^2 \exp \left[- \, \Phi_b(t)\right] \left[ \exp\left( q_e\,V(t)\right) - 1\right] ,
  \label{sc}
\end{equation} 
where $q_e$ is the elementary charge, $T$ the temperature and $A$ the Richardson constant \cite{sze_1981, mikheev_2014}. From Eq.(\ref{sc}) and employing 
Ohm's law, the resistance as a function of time was calculated as $R(t) = V(t) / I_s(t)$.

\begin{figure}[H]
\centering
\includegraphics[width=1 \textwidth]{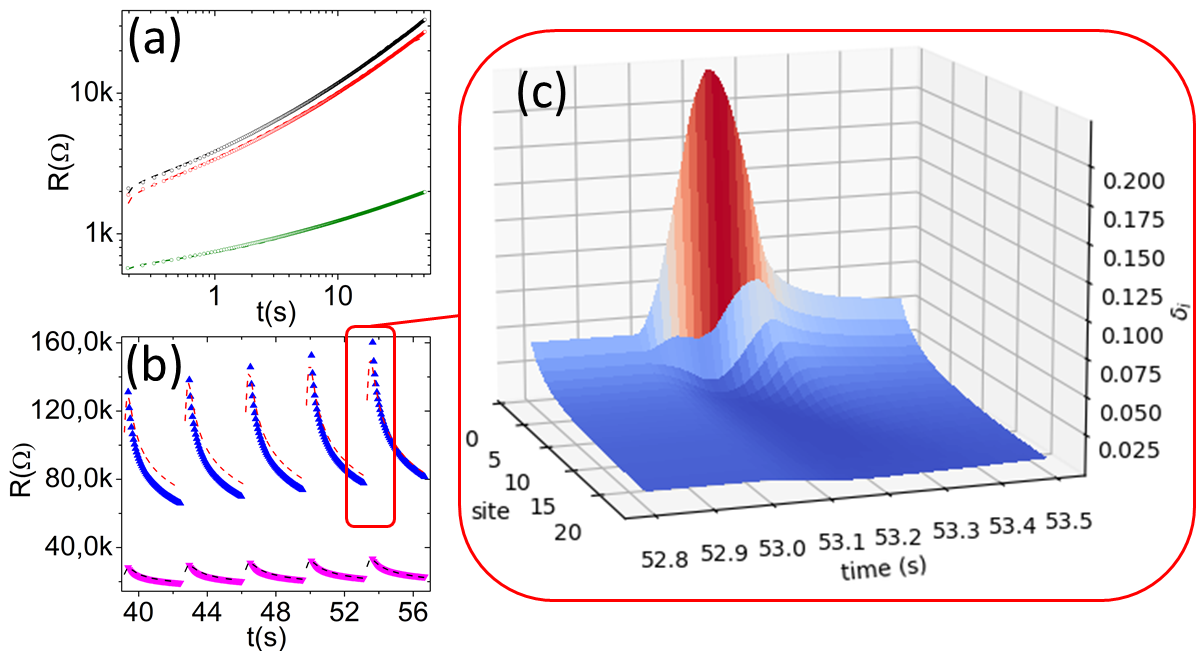}
\caption{ (a) Evolution of the resistance over time under different constant applied biases, starting from LRS (a set pulse of +2 V was used). Filled points correspond to experimental data for +50 mV (black), +100 mV (red) and +300 mV (green) and dashed lines to simulations. In each case, the evolution of the resistance follows a power law. 
(b) Time evolution of the resistance using pulsed protocol (as shown in Fig. \ref{Fig.result2}(c)) to obtain the double time scale behavior. We focus in the last five write/pulse events for -1 V (magenta triangles) and -1.5 V (blue triangles), for the reading bias of +300 mV. Filled points correspond to experimental data and dashed lines to simulations. Data corresponding to the negative voltage cycle of the protocol are not included. 
(c) Simulation of the time evolution of the fraction of OV per site for the time interval and resistance values indicated in (b) inside de red rectangle. See text for more details.}
\label{Fig.simulations}
\end{figure}

According to our model, the pure diffusion process (at zero external bias) reaches a steady  state where a significant fraction of OV accumulates
at the initial sites of the DR, which represent the interface between the metallic Co electrode and the STO. This OV accumulation is obtained 
regardless of whether the system starts from a LRS or a HRS 
(See  
Figs. (SM-2,SM-3) in  Supplementary Material). Since OV  carry a positive charge, they should be expelled from the DR under a sufficiently large positive bias which helps them to overcome the diffusion process. However, for relatively low external bias, one expects a competition between pure diffusion and drift induced by the bias. In order to check this, we measured the evolution of resistance over time of one device for three different constant reading biases (+50 mV, +100 mV and +300 mV), for around 15 s starting from the LRS (
following a +2 V set pulse).

In Fig. \ref{Fig.simulations}(a), we show the comparison between our simulated results for low bias, starting from LRS, and the experimental measurements, demonstrating excellent agreement. Our simulations indicate that 
when the system begins in the LRS — where the OV concentration is higher in the STO than in the DR — the dominant process in the absence of applied bias is  OV diffusion from the STO to the DR. However,  under a relatively low applied bias, OV drift counteracts diffusion, slowing it down and leading to a power-law increase in resistance over time. These results are in agreement with reported measurements in these devices at +300 mV showing that the evolution of resistance over time follows a power law, starting from the LRS \cite{Goossens_2023}.



Since we assumed a constant value for the electric permittivity $\epsilon$ in the model, fitting each experimental measurement for the three constant biases required adjusting the parameter $\beta = qn_T / \epsilon$ (which appears in the last term of Eq. (S3) in the Supplementary Material) and accounts for the contribution to the local electric field from the fraction of OV at each site.
The chosen values of $\beta$, shown in Table \ref{table_beta}, indicate that the permittivity must decrease as the applied voltage increases, a result consistent with reported experimental data for the electric permittivity of Nb:STO \cite{neville1972permittivity,yamamoto1998temperature}.

%

Having been able to reproduce the resistance changes over time for a constant bias, our next goal was to analyze 
and compare the resistance evolution under a pulsed protocol with our simulations. In Fig.\ref{Fig.simulations}(b) we show the last five cycles of the resistance measured with the pulsed protocol for writing voltage of -1 V (blue points) and -1.5 V (red points), read at +300 mV (already depicted in Fig. \ref{Fig.result2}(c)); together the simulated resistances (dashed line).
The simulations accurately capture the overall trend of the experimental data, although a transient response immediately after the writing voltage is noticeable.
It should also be noted that the last three cycles are better matched by the simulations, which is related to the choice of the constant value for the electric permittivity.

\begin{table}[]
\centering
\begin{tabular}{ | c | c | c | c | }
 \hline
 & +0.05 V [a.u]. & +0.1 V [a.u] & +0.3 V [a.u] \\ 
 \hline
 $\beta$ & 2.8 & 3 & 4 \\
 \hline
\end{tabular}
\caption{Values of $\beta= q n_T /\epsilon$ [a.u] obtained from the fits of the simulated curves to the experimental ones in Fig.(\ref{Fig.simulations}) (a) for the three voltages [a.u] employed in the simulations. }
\label{table_beta}
\end{table}

In addition in Fig.\ref{Fig.simulations}(c), we show the simulated time evolution of the fraction of OV per site for the time window indicated in Fig.\ref{Fig.simulations}(b). As expected from the previous analysis, during the application of a large negative bias, OV are pushed towards the metallic contact in the DR, while for a large positive bias, the OV are pushed out of the DR towards the STO. During the reading process (low applied voltage in the simulation), the drift/diffusion process previously described dictates the evolution of the resistance. 
To summarize, the competition between a small positive bias during the reading process and the diffusion of the OV towards the Co electrode in the DR gives rise to the modulation of the SB over time.

\section{Conclusion}

In this work, we investigated the synaptic plasticity behaviors in Co/Nb:STO Schottky memristive devices. Through experimental and theoretical analysis, we demonstrated how OV electromigration modulates the Schottky barrier height, enabling both LTP and STP -like effects. Our results reveal a power-law dependence of resistance changes on time during the reading process, which is indicative of a volatile memory effect. 
The gradual resistance decrease over time after the writing pulse is removed arises from the competition between OV diffusion and the residual electric field due to the reading voltage.
In addition to the volatile component,  we also observed a stepwise increase in resistance upon the successive application of writing pulses of different intensities. This response is attributed to a permanent redistribution of OV near the metal/oxide interface, which alters the Schottky barrier height in a persistent manner. Unlike the volatile response, which is dominated by OV drift and diffusion on shorter time scales, this behavior emerges when OV become deeply trapped at stable defect sites  within the depletion region.

To better understand these mechanisms, we developed and employed a theoretical model that captures the dynamics of OV migration under various pulsed and constant biases. This model successfully reproduced key experimental observations, including the modulation of resistance over time and the double time-scale behavior during pulsed protocols. The correlation between the OV distribution and Schottky barrier modulation underscores the importance of interfacial resistive switching mechanisms in these devices. The adjustments in the local electric permittivity parameter, consistent with experimental observations, further validated the robustness of the proposed model.

Our findings highlight the potential of Co/Nb:STO memristive devices for neuromorphic computing applications. The tunability of STP and LTP through pulse amplitude and frequency offers a versatile platform for emulating synaptic functionalities. This capability, combined with the device's inherent stability and multilevel resistive states, positions it as a promising candidate for real-time spatiotemporal signal processing and adaptive neural networks. Future work could focus on optimizing device architecture and exploring its integration into large-scale neuromorphic systems.

\section*{Supplementary Material}
See Supplementary Material for details on the model, computation of the energy potential profiles and for an analysis of the OV diffusion, together with the  associated profiles, in the absence of applied external stimuli.

\section*{Acknowledgments}
We acknowledge support from ANPCyT (PICT2019-0654, PICT2019-02781 and PICT2020A-00415) and EU-H2020-RISE project MELON (Grant
No. 872631). A. S. Goossens acknowledges financial support from CogniGron Centre. The experimental work was realized using NanoLab-NL facilities.

\section*{Authors Declarations}
\subsection*{Conflict of Interest}
The authors have no conflicts to disclose.

\subsection*{Author contributions}
WQ: Developed the theoretical model, performed electrical measurements and simulations. Writing–original draft, review and editing.
AG: Developed and fabricated the devices.  Writing–review and editing.
DR: Writing–review and editing. Funding acquisition.
TB:Writing–review and editing. Funding acquisition.
MJS: Model conceptualization. Writing–original draft, review and editing. Funding acquisition.
The results were discussed and their implications agreed upon by all authors. 

\subsection*{Data availability}
The data supporting this study will be
made available by the authors, without undue reservation.





\label{localE2}

\bibliography{reference}

\end{document}